\documentclass[12pt,preprint]{aastex}
\def\lapprox{\hbox{\lower .8ex\hbox{$\,\buildrel < \over\sim\,$}}}
\def\gapprox{\hbox{\lower .8ex\hbox{$\,\buildrel > \over\sim\,$}}}

\begin{document}

\title 
{\bf The Nova Stella and its Observers}

\bigskip

\author
{P. Ruiz--Lapuente \altaffilmark{*,**}}

\altaffiltext
{*}{ Department of Astronomy, University of Barcelona, Mart\'\i\ i Franqu\'es
1, E--08028 Barcelona, Spain. E--mail: pilar@am.ub.es}

\altaffiltext
{**}{Max--Planck--Institut f\"ur Astrophysik, Karl--Schwarzschild--Strasse 1,
D--85740 Garching, Federal Republic of Germany. E--mail:
pilar@MPA--Garching.MPG.DE}

\slugcomment{{\it Running title:} Nova stella 1572 AD}

\bigskip

\section{Introduction}

 In 1572, physicists were still living with the concepts of mechanics
 inherited from Aristotle and with the classification of the elements from 
 the Greek philosopher. The growing criticism towards 
 the Aristotelian views on the natural world could not turn very productive,
 because the necessary elements of calculus
 and the empirical tools to move into a better frame were still lacking.
 However, in astronomy, observers had witnessed by that time the comet 
 of 1556 and they would 
 witness a new one in 1577. Moreover, since 1543 when 
 {\it De Revolutionibus
 Orbium Caelestium} by Copernicus came to light, the apparent motion of the 
 planets had a radically different explanation from the ortodox geocentrism.
 The new heliocentric view slowly won supporters among astronomers.
 The explanations concerning the planetary motions would depart from 
 the ``common sense'' intuitions that had 
 placed the earth sitting still at the center of the planetary system. 

\smallskip

 Thus the ``nova stella'' in 1572 happened in the middle of the 
 geocentric--versus--heliocentric debate. The other ``nova'', the one 
 that would be observed in 1604 by Kepler, came just shortly after the
 posthumous publication of Tycho Brahe's {\it Astronomiae Instauratae 
 Progymnasmata}$^{1}$   which contains
 most of the debates in relation to the appearance of the ``new star''$^{2}$.

\smallskip 

 The machinery moving the planetary spheres, i.e. the
 intrincate system of solid spheres rotating by the motion
 imparted to them as the wheels in a clock, was not directly affected
 by the new heliocentric proposal. There, it was also needed to have
 a way to transmit movement. Most astronomers that accepted the heliocentric
 system stayed with the idea of solid spheres transmiting
 the movement, until the comets originating beyond the lunar realm and
 crossing therefore those ``spheres'' made their solidity impossible$^{3}$.
 
\smallskip

  What was then the scientific relevance of the event of 1572 and that of 
  1604? A first implication was to start to look at the stars 
  as material 
  bodies, at least some kind of stars. More generally, it gave rise to a 
  qualitative jump in the way to attempt the unification between microcosmos
  and macrocosmos. That can not be said to be 
  the obvious conclusion derived by
   most of the astronomers who observed and wrote
  about SN 1572, but by a few of them and by those of  
  later generations.
  Most XVI century astronomers were embedded in the
  Middle Ages notions of ``sympathies'' between the celestial and the 
  earthly realm. Amongst those who wrote about SN 1572 some were 
  practitioners of natural magic, as John Dee, actively involved in alchemy,
  and sharing the study of talismans and transmutation of metals 
  with that of mathematics and books of hermetic 
  and cabalist character.  
  The connection between microcosmos and macrocosmos was nearly that of
  the Middle Ages at the time of the advent of SN1572. The heliocentrist 
  astronomers who observed well in detail the {\it nova stella}, 
  Digges and Maestlin, read the relation of the planets and 
  their motions in terms of harmonies, both following the Neoplatonic
  views which inspired Copernicus. Sympathies 
  and harmonies were central concepts in the study of the cosmos
  for the astronomers 
  acquainted with the {\it Corpus hermeticum}$^{4}$ and the Neoplatonic
  writings. Along those lines, 
  the importance of geometrical beauty in the heavens
  is found in many XVI century astronomers.
   
  Staying far apart from the Neoplatonic influence, 
  Tycho Brahe would also keep the domain of the heavens away from that 
  of the earth. Natural tendencies of the stars and of the bodies in earth 
  would be different in both realms for him, suscribing here to Aristotle's 
  differentiation of movements. Tycho Brahe preserved as well the notion of    
  {\it quintessence} from the Greek philosopher, 
  the perfect immutable substance which permeates
  the heavens making them of different nature than the earth.  

  Therefore, either sympathies, harmonies or natural tendencies made the basis
  of the relation between the earth and    
  planets and stars. The ``nova stella'' gave the chance to rethink the
  difference between our realm and the one of the stars. Some of the
  observers, as Jer\'onimo Mu\~noz, were radical in the conclusions
  and looked for a non--hierarchical view of the cosmos, basically made of 
  matter stirring from the initial
  Chaos, and advanced mechanicist explanations. He wrote against the
  existence of quintessence as the perfect solid that made the sphere
  of the fixed stars$^{5}$. 

  In all the years following 1572, the practice of printing and
 translating books helped to spread the ideas of the various
 observers of the {\it nova stella}. And most of the views
 and records by astronomers and philosophers 
 ended up in the  volumes of Tycho Brahe's {\it Progymnasmata}. 
 Though Tycho Brahe was not keen in abandoning 
 the idea of the solidity of the celestial spheres and the  
 immutability in the sphere of the fixed stars, he took care of
 quoting the debates and most of the observational facts. Systematically
 noting the evolution in brightness of the event,  his account in 
 the {\it Progymnasmata} gives us the place  
 where the {\it nova stella}  should be located within the
 class of stellar explosions to which it  belongs$^{6}$.

\bigskip

\section{The records in modern perspective}

The observers of the {\it nova stella}, known nowadays as SN 1572 or
Tycho Brahe's supernova, contributed to the empirical knowledge of
this event by measuring its position and 
recording the brightness until it faded. 

The observers measured angular distances from the 
supernova to the stars in Cassiopeia by naked eye and with the help
of threads, rulers or sextants. They were limited by the eye's 
 precision in measuring angular distances, which is about 1 arcmin, and 
by the inaccuracies of their methods.
Brahe, Digges and Mu\~noz achieved the 1 arcmin precision of the eye in 
their measurements of the angular distance between the stars 
in Cassiopeia and the supernova. Tycho Brahe was here as well the most 
exhaustive of them, measuring angular distances to a larger number of stars.
Mu\~noz got to the limiting accuracy of the eye
 when obtaining the distance between the nova and 
 $\alpha$ Cas and $\beta$ Cas while doing an error of 10 arcmin in the
distance between the new star and  $\gamma$ Cas. The same is found 
in Digges and Brahe for some of their measurements. 
Hagecius, in contact with Mu\~noz, learnt from him 
how to correct his results. His original 
measurements were up to 20 arcmin off$^{7}$. Stephenson
and Clark$^{8}$ reconstructed the historical 
position of the {\it nova stella} from 
Tycho Brahe's measurements giving a final standard deviation of 22 arcsec. 
They found in the writings by Tycho Brahe the likely reason of the
uneven quality of his measurements, as he did not use always his 
half--sextant. Eliminating measurements 
that were severely off, and using Digges observations, the original
position by Tycho Brahe, well north of
the centroid of the remnant, comes into agreement 
with the positions of the rest of observers
who are closer in declination to the centroid. 
Both Mu\~noz and Digges got 
the declination right while carrying a larger error in right ascension.
Green has recently reexamined 
the positions given by those observers and for others
whose work had not been examined so far$^{9}$. 
Differences in the historical position of the {\it nova} 
as large as the radius of the present remnant are found.  
Overall, the accuracies of the observers are very different, as some 
of them used very crude methods.

While the position of the supernova can not be retrieved at an accurate
level from the measurements done in 1572, the records of the brightness
have turned out to be very useful. Just a paragraph in the otherwise rather
extended account in the {\it Progymnasmata} by Tycho Brahe 
helps us to build up the visual {\it light curve}, i.e. the
evolution of luminosity with time for this event. 
Tycho Brahe did not quote the specific date of the observations
but the month or even and interval within two months. However,
for most of what one can extract, such an interval is enough. 
He and the rest of astronomers compare the magnitude of the supernova
to that of stars and planets. 
Mu\~noz provides the most constraining 
account on when the explosion took place as he first observed it on Nov 2
but saw nothing$^{10}$. He writes 
that he was certain that on 
November 2, 1572 the ``comet'' was not in the sky, as he was
``teaching to know the stars to a numerous group of people that
evening''. In Nov 11 it was already seen by some shepherds in Ontiniente
(near Valencia), and  the author of the ``Book of the New Comet'' 
seems to have worked several chapters by Jan 7 1573, as 
he writes: ``when it began to be visible, it looked larger
than Jupiter, and now, on January 7, 1573, it already looks smaller
than Jupiter; this could have happened because it had risen higher 
than where it was when it first appeared''. 
This note is an upper limit to the
rate at which the luminosity of the supernova declines.  
It is consistent with the account given by Tycho Brahe that 
on January it was a little fainter than Jupiter and brighter than
the brightest stars of first magnitude. Tycho Brahe first noticed the
supernova on Nov 11 and finished recording the brightness on March 1574.
On Nov 11 and Nov 16 it was observed as well by Caspar Peucer and 
Johannes Pr\"atorius, whose records help to draw the premaximum
phase in the evolution in luminosity of the supernova. 
From the combined records one can reconstruct
the light curve of this event and say where it stands within this class 
of explosions.  
From the color evolution one can also state that Tycho Brahe's nova
was in fact a normal thermonuclear supernova as those used today in cosmology
as distance indicators. The observations  by Brahe, Maestlin, Mu\~noz and
Pr\"atorius  were the most useful to trace
 the evolution in brigthness and color$^{11}$. 

If we consider the errors that we can safely assign to those
determinations, no doubt that the light curve observations should be 
regarded as precise. They were at the frontier of what it was
possible to do before the telescope came into use.  
The method used by those authors, specially by Tycho Brahe,
 of noting when it appeared and 
how it changed in brightness, is the same one used
in our times. Even the observational limit that they reach of 
0.2 mag is in accuracy at the limiting magnitudes for the eye$^{12}$.

\section{The {\it nova stella} and the turn of the cosmos}

Since the appearance of the latin translation of the {\it Corpus 
hermeticum} by Marsilio Ficino, in 1463, 
a growing interest in these
texts as containing fundamental truths of the Christian religion 
spread through Europe. The ancient cult to the sun by the Egyptians, 
viewed through the 
Christian prism, had an influence in Copernicus and his heliocentric
 system$^{13}$. Astronomers 
of Neopythagorean and Christian hermetic views such as Maestlin 
were all reluctant to accept SN 1572 as a 
natural phenomenon$^{14}$. They considered it a star that appeared
as a supernatural act. That was as well the position that Maestlin's 
disciple, Kepler, had towards the {\it nova} of 1604. 
It was also the position of Tycho Brahe regarding SN1572.  

In the most conservative realm, some 
observers considered that the star of 1572  had allways been there
but a change in the air or in the celestial aether allowed it to be seen. 
Of that opinion was Francisco Valles, who wrote a treatise   
criticised in Tycho Brahe's {\it Progymnasmata}$^{15}$. 
Francisco Valles  had acknowledged three ways to know 
the truth: experience, reason applied to the experience and authority
of the ancient and scholars, the third of which being the less suitable. 
However, regarding the {\it nova}, he would not depart
from tradition. 

Tycho Brahe himself defended as well traditional views for many
years and did not change his position,
 motivated by the astronomical events that
happened in the decade 1570-1580, until much later on. For long
he denied that comets or meteors
could be among the planets and the fixed stars and believed in the 
solidity of the celestial spheres, as it has been 
documented in letters to Caspar Peucer and Rothmann. 
He and Caspar Peucer held correspondence
on the solidity of the spheres still around 1588, well 
after SN 1572 and the comet of 1577.  Only gradually he
accepted that the solidity of the spheres did not fit the observations
and moved to admit a liquid nature for those. In all this, he seems to
have tried to look into the Scriptures as a guide to find the true
cosmological picture. The physical way to move the heavens was lacking
if the spheres were not solid and assembled as in clockwork. 

\noindent
He finally conceded to move away from previous views$^{16}$: 

\noindent
\begin{quotation}
 {\sf 
\noindent
  Heaven is not made up of real, durable and impervious orbs to which
  the stars are affixed and travel, but it consists of a substance
  that is very clear, very thin and very fine. This makes the courses 
  of the seven planets free so that they move without any slowing 
  wherever their natural impetus and their knowledge carry them. This
  was not seen by the ancients or even the greater part of the
  moderns, nor even conceded because it was never doubted. For it is
  enough for the restoration of astronomy to admit it as settled and 
  known.} 
\end{quotation}

Lacking other better explanation, 
when the clockwork shell machinery transmitting the movement 
disappeared, the planets somehow would be following a natural impetus
 or knowledge, according to Tycho Brahe. 
  
On the Neoplatonist shores, the harmonies and beauty necessary 
to the motions in
heavens were confronted with the choice of an acceptable level of  
 departure from the aimed circular perfection. 
In the world of ``sympathies'' or ``consonances''
 as ruling the planetary motions, 
exceptions were sometimes judged  to add to the beauty of the cosmos:  
John Dee$^{17}$  would consider that planets 
move according to law of ``consonance'' or
``sympathy'', though some ``dissonance'' or ``antypathies'' are
seen in the retrograde movement of planets
contributing to the ornament of the universe. The degree of tolerance
to exception within the tradition of natural magic could be very  
large. 

In general, the Neoplatonist authors preserved Aristotelian notions
of movement. Digges 
in his {\it A perfit description of the Caellestial Orbes}$^{18}$ combines
them with Platonic influences.

Adding to their common ground, Digges and Brahe 
kept the distinction between noble and less
noble movement as a separation of the motions between the realm 
of the terrestrial and the heavens and refer to ``natural tendency''
for the motions. 
In Digges, the domain of movements is in accordance
to the noble or less noble realm that characterized  as
well the composition of the bodies, that of the harmonious as in
the heaven or the forced or unnatural in our earthly realm. 
He distinguishes two types of motions, straigth or circular, and 
discusses whether a violent movement or a natural movement
is found in the earth's motion.

 The distinction between the domain of the perfect
 and the domain of the corruption and the perpetuation of the system of four
 elements (air, earth, fire, water) as the basis of the material world
 is the referent into which astronomers would state their views. In Tycho
 Brahe,
 whether quintessence was a cristal or a metal had been addressed as well.

  Though all the ancient cosmogonies had a sequence of arrangements of planets
  which was more or less fortunate, transforming such sequence into a 
  system of solid shells fitting inside each other resulted in the enduring
  basis of celestial mechanics. 
  The solidity in the spheres is no Aristotle's invention since  
  solid wheels whose circular openings would make the planets 
  are proposed in Anaximander. Previously to Aristotle, Plato had revived
  the animistic view of the cosmos
  in his {\it Timaeus}, where the world is argued to be
  a sphere in analogy with 
  the man, as it is the best shape to be given to it as a living
  organism that does not need arms nor legs.  Aristotle continued with
  the animistic trend from Plato's {\it Timaeus}. Though in Aristotle 
  the system is more elaborated and the intelligences moving the cosmos are
  eternal: the Universe is without beginning. 
  
\noindent
``In this way that according to Aristotle, the heaven and its intelligences
are so eternal as the first cause, and over them there is no more power
than the first mover, being in the supreme region makes the turn of
the heaven and makes the intelligences to move and with that movement
to place to move the 
inferior orbs'', summarizes Mu\~noz in his ``Book of the New
Comet''where he points the need to abandon this idea.
Along the chapters
of the book, he follows  very orderly the various
aspects that this new event raised and derives far--reaching consequences
 for cosmology. From the lack of 
parallax of this ``comet'' Mu\~noz deduced that it did not happen in
the ``air'' as in the comet theory of Aristotle and that there could
be mutability in the heavens and that those were not solid. 
The comet appeared in 1556 had probably helped to build
his opinion on the continuity of the heavenly and earthly domains. 
In the account on the {\it new comet}
 he clearly acknowledges the possibility of combustion in the
heavens and that they are made of the same elements as the sublunar world.
A unified 
cosmos where physical changes occur, and where motion is transmitted through
rotation from the initial Chaos as suggested by Anaxagoras:

\smallskip

\noindent
\begin{quotation}
\noindent
{\sf Aristotle understood that the comets were done in the superior
region of the air and not in the heavens, as Democritus and Anaxagoras
(philosopher and great mathematician) want, whose opinions Aristotle
as he was not an astronomer did not understand, neither has the 
base to understand the Caldean and Egyptian doctrines$^{19}$.

  ...

\noindent
I have many experiences that force me to be in regard to
the comets and other opinions closer to Democritus and Anaxagoras (who
was philosopher and great astrologer) than to Aristotle, who in his 
works does not show to know astrology, while admiring the curiosity and
diligence of the astrologers, and Egyptian, Caldean and Babilonic 
priests$^{20}$.}
\end{quotation}

Mu\~noz calls comet to the nova  meaning a new kind of comet, in the region 
of the fixed stars, which shows 
no parallax  and does not cross transversally
the skies, though likely moves just approaching
and receding along our line of sight.   
He thinks that comets  are bodies 
similar to the planets, but not planets themselves, and  that 
they are made near the 
poles because the heavens are denser there due to
 the constellations. One has to keep in mind that Jer\'onimo Mu\~noz
noticed that the {\it comet} was aligned with the Milky Way,
 and thought that it likely arises from it. When it
comes to say how do these bodies get their motion, he looks for light
as causing the effect, i.e.  
the planets, with their rays gathered towards the otherwise
cold regions of the constellations would lighten a part of those
regions, and from their rays and the stars a comet, or fire is
made that ``gathers strengths and circular movements different from 
those of its parents, or makers, and sometimes similar to them''.
For the same reason, the conjuction of planets would favor the production
of comets.  

Moreover, reports on a variety of ``comets''  might have helped
in making the supernova to be classified in such a way and not
as a star. According to Mu\~noz, there were records of various types
of comets and he devotes a chapter to their classification. The Comet
of 1572 did not have parallax and was in the region of the fixed stars. 
But that would not be the first time that such 
kind was seen, as there are comets that show no
parallax: 

\noindent
\begin{quotation}
{\sf
\noindent
 In no author I find Comet like that, who resembles more a
star than a Comet. Ptolemy and Pliny mentioned some 
Comets that do not move and do not get apart
from the stars from which they came out, but they do not describe
their form. Lucanus says that before the civil war between Julius Caesar
and Pompeius, {\it ignota obscurae viderunt sidera noctes}, that were
Comets in the way of stars, not known to scholars: from that
genre of Comets is without doubt ours$^{21}$}. 
\end{quotation}

Other than in the theory of comets, 
the attraction of this author to early Greek philosophers, and in
particular to Anaxagoras, seems to be manifold. One aspect is 
the attraction to the ``nous'', Mind that organizes the cosmos.
 Anaxagoras and his ``$\nu o\hat{\upsilon}\varsigma$''  are at the start of
 rationalism, but it is also in the Genesis$^{22}$.
 Another link with the Presocratic philosopher is  
the rejection of the Aristotelian theory of the natural place
of the elements in the cosmos while holding a view of matter made everywhere 
of the same elements. This view brings him close to 
the ``seeds'' of Anaxagoras or ``atoms'' of 
Democritus as the basis of the material world. 
 The  Aristotelian view of matter seems a step back if we consider 
  the intuitions of a unifying principle under the diversity of
  the material world of some early Greek philosophers. 
Mu\~noz goes back to
the philosopher Anaxagoras, who held the opinion that
no thing is born or perishes, but composes itself and disolves 
itself from the existent things. Birth would be composition and
perishment disolution. Parts of the script deal with 
the play of those mixtures of elementary components
 in the heavenly bodies and a chapter is entitled 
``Heavens and stars are not quintessence, but 
that they have a debt and relation with the elements''. 

\smallskip

The origin of the cosmos
from Chaos and the whirlpool model of transmision of movement is 
another connection that would place Mu\~noz in the line of thought 
of turbulence in the heavens and cosmic continuity, 
that one leading to Descartes's
planetary model.
Rotation as inherited from the original Chaos,
is noted by him to be proposed by Anaxagoras (and also found in
Middle East cosmogonies).   
The motion, the rotation of the heavens, would have
been there from the original moment of Chaos and since then it would
have been preserved and passed on to the various bodies, i.e. rocks, in which
the matter would be stirred. 
The Anaxagoras {\it nous}, the principle 
responsible for the material realm,  governed the
 rotation at the beginning, starting on a small area to
end up in a larger one and a still larger one in the future. The {\it nous}
knows all the things mixed, separated and divided. The same rotation
that was set on the things made them split. Within this view which is 
mechanicist after setting up the rotation, it is
clear that there is no need for the existence of solid spheres passing
movement. Moreover, for philosophers acquainted with the Middle
East cosmogonies, Anaxagoras, rather than Aristotle, would be the philosopher
that best explain the views transmitted in the
Bible, Babilonic, and Middle East cultures to the ancient Greeks. 
Along these lines, Jer\'onimo Mu\~noz has in mind a Universe 
opposite to the eternal cosmos of Aristotle, i.e. 
the cosmos as order coming from  chaos as in the
Scriptures$^{23}$.

Thaddeaus Hagecius, who had contacted Mu\~noz on the subject of the nova,
would refer largely  to the Presocratics in his treatise published in 1574.
 This observer of SN 1572 had, however, interests in the other 
lines of thought of the time$^{24}$. He widely discussed on
the Cabala with John Dee and was in touch with Tycho Brahe. 

The mechanicist picture of the ``Book of the New Comet'' 
 was very different from the
point of view of the Neopythagorean--Neoplatonic astronomers
Digges, M\"astlin and others (tradition culminated by M\"astlin's 
disciple Kepler). Their view, other than description in terms of
purity of geometrical forms in the heavens, would not refer to
transmission of impetus but, in the case where this is addressed, to
a possible magnetic influence of the central body (the Sun) on the
planets around. In Kepler, this idea is well developed up to the point
that he considers the planets as possibly  dragged by the magnetic
force arising from the Sun. 

Kepler would propose as a force to hold the planets$^{25}$:

\noindent
\begin{quotation}
{\sf 
\noindent 
 This force which takes hold of the planetary bodies and
transports them is an incorporeal emanation from the force which is
located in the sun. 

...

\noindent
The force which extends out from the sun to the planets moves them in
a circle around the immovable body of the sun.

...

\noindent
We may believe that in the sun there is no force atracting the
planets, as in a magnet (for they would continue to approach the sun
until they were completely joined to it), but only a directional
force. Hence the sun has circular fibers which sweep around in the
direction shown by the zodiac. Therefore the perpetual rotation of the
sun is accompanied by the circular rotation of that moving force or outflow
of the emanation from the sun's magnetic fibers. This outflow is
diffused throughout all the planetary distances, and its rotation
occurs in the same period as the sun's}. 
\end{quotation}

Along with Thomas Digges, Michael Maestlin 
was the other Copernican observer of the {\it nova stella} 
who significantly contributed with observations. His brightness 
and color estimates are useful accounts to retrieve the light curve 
of the supernova. Maestlin interests in the hermetic
tradition and the Christian Cabala were as well as his Copernicanism
an impacting influence on his disciple Kepler.

\section{The {\it nova stella} and the Pythagorean views}

\smallskip

Digges, in his treatise ``A perfit description
of the celestial orbes according to the most ancient doctrines  of
the pythagoreans'' follows still
the line of distinction between movements in the celestial spheres and
those in the sublunar realm, and shows an aesthetic preference for
the Copernican model. In fact, all what is contained
in Digges's treatise follows Copernicus's {\it De Revolutionibus
Orbium Caelestium}, as Thomas Digges's
 intention was to improve the edition of the existing book by his father
 Leonard Digges appending what he considered the right cosmological model.
Thomas Digges is convinced of Copernican  heliocentrism.

\noindent
His view on motions keeps Aristotelian concepts but outlines the geometrical
relevance of the tendencies in the heavens. 

\noindent
\begin{quotation}
{\sf
\noindent
Tight or straight motion only happen to those things
that stray and wander, or by any means are thrown out of their natural
place. But nothing can be more repugnant to the form and ordinance of
the word, than that things naturally should be out of their natural
place. This kind of motion therefore, that is by right line, is only 
accident to those things that are not in their right state or 
natural perfection, while parts are disjoined from the whole body, and
convey to return to their unity therefore again$^{26}$.}
\end{quotation}

  Along these lines he defines gravity as a natural tendency of
  the bodies: 

\noindent
\begin{quotation}
  {\sf 
\noindent
 Gravity is not anything but the natural tendency given by the 
   divine providence of the Creator to the parts, for which virtue
  they tend to unite with the bulk and to restaure in this way the
  unity or integrity under the spherical shape. It is very likely that 
  this same property or affection is on the Moon and the other noble
  celestial bodies, in the way that they aim to gather their parts
  and preserve the spherical figure$^{27}$}. 

\end{quotation}

Thus Digges and many
  others  would see in the spherical shape the natural one 
  for the celestial bodies, so very much as in the Pythagorean tradition
  and that of the {\it Timaeus}. 

Bodies tend to keep approaching 
 following a sphere. Circular, spherical shape in the realm of the
the noble bodies would prevail and there is an aesthetic preference for the
sun at the centre of the solar system.
As in Copernicus, Digges writes that Hermes Trismegistus called the sun the 
visible God and king of the planets.

Amongst the most appreciated contributions by Digges given in his treatise
are the arguments on why the motion of the earth will not stir it
apart: ``But anyone who maintains the earth's mobility may say
that this motion is not violent, but natural''$^{28}$. He compares it with 
sailing in a smooth sea: ``A ship 
carried in a smooth sea does move so steadily that all
things on the shores and the seas seem to the sailors to move while
themselves remain at rest together with all things that are aboard
with them. So, surely it may happen that the earth, its motion being
natural and not forced, but most uniform and
unperceivable, moves in such a way that to us, who are sailing therein, 
the whole world may seem to roll about''$^{29}$.

 Digges uses arguments in favor of an infinite Universe, amongst them
  the suggestion that the heavens could be infinite
  because it makes no sense that {\it nothing} will restrain the
  cosmos:

\noindent
\begin{quotation}
  {\sf
\noindent
   without the Heaven there is no body, no place, no emptiness;
  no, not any thing at all, whether heaven should or could farther
  extend. But this surely is very strange that nothing should have
  such efficient power to restrain something, the same having a very 
  essence and being.$^{30}$} 
\end{quotation}

  This argument is Copernican$^{31}$, as much as the views on gravity
 and the motion of the earth, 
 but was stressed in Digges's treatise including a diagramme
 of the cosmos where the sphere of the fixed stars extends to 
 infinity.

\section{Geometry and the unified cosmos}

  Though acknowledging changes in the heavens, the picture
  of geometry ruling the cosmos prevailed among most
  astronomers of this generation that shared the Copernican
  influences and read the mind of God in geometry.
   Digges and other Neopythagoreans  did
  not give up the spheres in view of the {\it nova stella}. The perfection
  beyond the sphere of the
  Moon would still be the atmosphere where Kepler developed his studies. 
  The geometrical approach would lead in Johannes Kepler to 
  the description of the movements of the planets that inagurated a new
  epoch in science. Isaac Newton, breathing similar philosophical influences,
  would much later bring the mathematical baggage for understanding
  Kepler's laws and the motion of the bodies. With him 
  gravity would move from the philosophical discourse 
  into the discourse
  of science. But that would happen one century later.

  The line of thought leading to the unified cosmos and the 
  mechanicist transmission of impetus in it is already present in the
  debate around 1572, as has been shown here. It 
  would gain a mathematical formulation in Descartes and his
  vortex cosmology in {\it The World or  Treatise of Light}. At the 
  beginning of the XVII century atomism would be recovered. The corpuscular
  view of the cosmos common to earth and planets would be the basis of 
  mechanics and of the discussion about gravity. That current of thought is 
  also the seed of the Newtonian mechanics$^{32}$.

  Those steps occured in the century following the ``nova''. 
  For what concerns the  ``nova'' relevance, 
  it certainly implied a gain of the 
  importance of the empirical realm in relation
  to the philosophical discourse used for centuries in cosmology.
  The observers witnessed mutation in the heavens, and such 
  mutability held a key to the relation between heavens and earth. 
  It would take, however,
  five centuries more to start to  know the closeness of the 
  two realms. 
  
  The observations of the {\it nova stella} set up very high standards
  in the systematic way to proceed in astronomy. 
  Tycho Brahe was precise and complete about 
  the observations and his records would be 
  discussed for the next five centuries.

\section{\bf Notes and references} 

\noindent
$^{1}$Tycho Brahe, {\it Astronomiae Instauratae Progymnasmata} 
(Uraniborg, 1602), in
Tycho Brahe Opera Omnia, 15 vols,
 ed. J.L.E. Dreyer (Amsterdam: Swets \& Zeitlinger, 
1972), vols II and III.

\noindent
$^{2}$Kepler, who took care of publishing the work by Tycho Brahe, 
brought to light his ``New Astronomy'' in 1609, five years after seeing what
he considered a sign. He completed his first law of planetary 
motion one year after his {\it nova}. 

\noindent
$^{3}$ It is indeed diverse when and why the heliocentrist astronomers abandon
the solidity of the orbits. 
Kepler first acknowledges that the celestial substance is tenous 
in his Optics written in 1604, well after the comets. 

\noindent
$^{4}$ References to 
 Hermes Trismegistus are common in the writings of the observers
 of SN 1572. All the observers mentioned here were acquainted with the
 {\it Corpus Hermeticum}. 
 For an English translation, see {\it Hermetica: The Greek Corpus
 Hermeticum and the Latin Asclepius}, with Notes and Introduction by
 Brian P. Copenhaver (Cambridge: Cambridge University Press, 1995). 

\noindent
$^{5}$ Jer\'onimo Mu\~noz {\it Libro del Nuevo Cometa} (Valencia: 
Pedro de Huete, 1573), re--ed. with introduction, appendices and 
Mu\~noz's letters by V. Navarro--Brotons (Valencia: Hispaniae
Scientia, 1981), hereafter cited as LNC. 

\noindent
$^{6}$See W. Baade, 
 {B Cassiopeiae as a Supernova of Type I}, 
{\it Astrophysical Journal} {\bf 102} (1945), 309--317; 
P. Ruiz--Lapuente, {Tycho Brahe's Supernova: Light from Centuries Past},   
{\it Astrophysical Journal}, 612 (2004), 357--363.
Walter Baade was the first to find that the {\it nova} was in fact
a supernova of Type I. That class of supernovae has been regrouped among
supernovae of core--collapse (encompassing 
Type II and a part of the Type I class, i.e. 
Type Ib and Type Ic)  and thermonuclear supernovae (Type Ia), those in which
no collapsed object is formed but a white dwarf is explosively disrupted.
It has been discussed for long whether SN 1572 was peculiar. Through 
the retrieval of the records and by measuring the extinction towards the
supernova it can be seen that it was the most common type of event, such
as those used for distance estimates to other galaxies. The rate 
of decline is found to correspond to a ``stretch factor''
$s$ $=$ 0.9 $\pm$ 0.05. 

\noindent
$^{7}$ This is seen in the letters 
by Mu\~noz. The correspondence between Hagecius and 
Mu\~noz is included in the letter of Mu\~noz to Reisacher.
Mu\~noz to Reisacher LNC, 107.

\noindent
$^{8}$ F. R. Stephenson \& D. H. Clark {\it The Location of the Supernova 
of AD 1572}, Q.J.R.A.S.,{\bf 18}, (1977), 340--350. The positions are
given in the equinox of 1950. They would correspond to   
an uncorrected position in right ascension and declination in the
equinox of 2000.0 of 
RA= 00$^{h}$ 24$^{m}$ 48$^{s}$  and Dec=64$^{0}$ 08$^{'}$ 49$^{'}$ (J2000.0) 
and a reconstructed position of RA=00$^{h}$ 25$^{m}$ 20$^{s}$  
and Dec=64$^{0}$ 07$^{'}$ 55$^{''}$ (J2000.0). 

\noindent
$^{9}$ D. W. E. Green, {Astrometry of the 1572 supernova (B Cassiopeiae)},
{\it Astron. Nachr.} {\bf 325}, 9 (2004), 1--13. The astrometry of SN1572 
from most observers of SN 1572 can be found here.

\noindent
$^{10}$ See Mu\~noz, LNC, 7.

\noindent
$^{11}$ 
In Ruiz--Lapuente, 2004, 360-361 a reconstruction of 
the supernova visual light curve and color 
light curves is found. It is derived where SN1572 stands amongst Type Ia
supernovae in intrinsic luminosity and rate of decline. 
It is also compared to other Type Ia supernovae of the last decade.  

\noindent
$^{12}$ 
In today's astronomy, using CCD detectors one quotes typical accuracies of 
0.01 mag in these supernovae when they happen in other galaxies. 

\noindent
$^{13}$ 
Thomas S. Kuhn, {\it The Copernican Revolution}, (Cambridge: Harvard 
University Press, 1957), 130--131.  

\noindent
$^{14}$  M. A. Granada discusses the interpretation of the {\it stella nova}
as a supernatural phenomenon in
{\it El Umbral de la Modernidad}, (Barcelona: Herder, 2000), 397--403. 

\noindent
$^{15}$ Tycho Brahe Opera Omnia, III, 87--92.

\noindent
$^{16}$ See Brahe to Peucer, undated letter of 1590, Tycho Brahe Opera
Omnia, VII, 231,
 Brahe to Peucer, 21 February 1589; VI, 177; 
 Howell, K. J. 1998, The Role of Biblical Interpretation in
  the Cosmology of Tycho Brahe, {\it Stud. Hist. Phil. Sci.}, 
 29, 4,  515--537; Granada, M. A. 2004, Astronomy and Cosmology 
 in Kassel: the Contributions of Christoph Rothmann and his relationship 
 to Tycho Brahe and Jean Pena, in {\it Science in Contact at the 
 Beginning of the Scientific Revolution}, (Prague: Jitka Zamrzlov\'a, 2004),
 236--249.

\noindent
$^{17}$ 
John Dee 1567, {\it Propaedeumata Aphoristica} 
in {\it John Dee on Astronomy, Propaedeumata Aphoristica},
edited and translated with notes by Wayne Shumaker,
(Berkeley: University of California Press, 1978). 

\noindent
$^{18}$ {\it A Perfit Description of the Celestiall Orbs according to the
most auncient doctrine of the Pythagoreans}, Thomas Digges (1576), Folio 46
in the volume including Leornard Digges treatise
{\it A Prognostication Everlastinge, Corrected and Augmented by Thomas
 Digges} (London: Thomas Marthe, 1576), re--edited in 
{\it The English Experience}, (Amsterdam: W. J. Johnson, Inc., 1975), vol. 727,
hereafter PDCO.

\noindent
$^{19-21}$ Mu\~noz, LNC, citations from p. 2, 26, 15 respectively.

\noindent
$^{22}$ Walter Burkert in ``From Homer to the Magii'' points
out a paralellism between the cosmological 
picture in Anaxagoras and the Genesis, {\it Da Omero ai Magi} 
(Venezia: Marsilio Editori, 1999). 

\noindent
$^{23}$ Mu\~noz, LNC, 25. 

\noindent
$^{24}$ The works written by Hagecius are analysed in Granada, 434--438.

\noindent
$^{25}$ Johannes Kepler {\it Dissertatio cum nuncio sidereo}.  
 {\it Conversation with Galileo's Siderial Messenger}
translated with an introduction and notes by Edward Rosen in
 {\it Kepler's Conversation with Galileo's Siderial Messenger}
(New York: Johnson Reprint Corp. 1965), 42 and note therein.

\noindent
$^{26-30}$ Digges, PDCO, citations in p. 12, 13, 9, 11, 10
respectively from the reproduced manuscript. The quotations from 
Digges's English manuscript are here rephrased 
for the sake of text uniformity. 

\noindent
$^{31}$ Digges's citations $^{26-30}$ are similar to excerpts from
 N. Copernicus's chapters VIII and IX of 
Book I {\it De Revolutionibus Orbium Caelestium}, translated with 
 an introduction and notes by  A.M. Duncan in 
{\it On the Revolution of the Heavenly Spheres} (ed. David \& Charles, 1976). 

\noindent
$^{32}$ For Descartes influence in Newton and the
path towards the laws of motion, see Kuhn, 1957, 258--259.  

\end{document}